\documentclass[10pt,letterpaper]{article}
\usepackage{opex3}
\usepackage{color}
\usepackage{SIunits}
\usepackage{epstopdf}
\begin{document}

%%%%%%%%%%%%%%%%%% title page information %%%%%%%%%%%%%%%%%%
\title{Ptychographic hyperspectral spectromicroscopy with an extreme ultraviolet high harmonic comb}

\author{Bosheng Zhang$^*$, Dennis F. Gardner, Matthew H. Seaberg, Elisabeth R. Shanblatt, Christina L. Porter, Robert Karl, Jr., Christopher A. Mancuso, Henry C. Kapteyn, Margaret M. Murnane and Daniel E. Adams}

\address{JILA, University of Colorado, 440 UCB, Boulder, Colorado 80309, USA}

\email{$^*$Bosheng.Zhang@colorado.edu} %% email address is required

% \homepage{http:...} %% author's URL, if desired

%%%%%%%%%%%%%%%%%%% abstract and OCIS codes %%%%%%%%%%%%%%%%
%% [use \begin{abstract*}...\end{abstract*} if exempt from copyright]

\begin{abstract}
    We demonstrate a new scheme of spectromicroscopy in the extreme ultraviolet (EUV) spectral range, where the spectral response of the sample at different wavelengths is imaged simultaneously. It is enabled by applying ptychographical information multiplexing (PIM) to a tabletop EUV source based on high harmonic generation, where four spectrally narrow harmonics near 30 nm form a spectral comb structure. Extending PIM from previously demonstrated visible wavelengths to the EUV/X-ray wavelengths promises much higher spatial resolution and more powerful spectral contrast mechanism, making PIM an attractive spectromicroscopy method in both the microscopy and the spectroscopy aspects. Besides the sample, the multicolor EUV beam is also imaged in situ, making our method a powerful beam characterization technique. No hardware is used to separate or narrow down the wavelengths, leading to efficient use of the EUV radiation. 
\end{abstract}

 % REPLACE WITH CORRECT OCIS CODES FOR YOUR ARTICLE, MINIMUM OF TWO; Avoid using the OCIS codes for “General” or “General science” whenever possible.

%%%%%%%%%%%%%%%%%%%%%%% References %%%%%%%%%%%%%%%%%%%%%%%%%

%%%%%%%%%%%%%%%%%%%%%%%%%%  body  %%%%%%%%%%%%%%%%%%%%%%%%%%
\section{Introduction}
New imaging capabilities are driving revolutionary advances in science and technology, making it possible to image biological systems in three dimensions with stunning detail. For materials systems that cannot be labeled using fluorescent markers, and that are opaque to visible light, shorter wavelength light has advantages. As nanosystems relevant to the semiconductor industry become increasingly complex, there is a critical need for microscopy techniques that can distinguish between different materials. This capability is realized using spectromicroscopy techniques in the extreme ultraviolet (EUV) and X-ray range, which take advantage of element-specific absorption contrast. However, established techniques such as near edge X-ray absorption fine structure (NEXAFS) \cite{Ade1993}, X-ray absorption near edge structure (XANES) \cite{Ade1990}, and X-ray fluorescence \cite{Horowitz1972} spectromicroscopies can only record images at one wavelength at a time: either a monochromator or spectrometer must be scanned in order to collect a full series of images. In contrast, ptychographical information multiplexing \cite{Batey2014} (PIM), which is an extension to ptychographical coherent diffractive imaging (CDI), enables the collection of images at multiple wavelengths simultaneously. When combined with EUV and X-ray wavelengths, this technique can be used to produce images with elemental contrast without the need for wavelength scanning.

Ptychographic CDI \cite{Faulkner2004, Rodenburg2007, Thibault2008, Seaberg2014, Zhang2015} has achieved great success due to its high fidelity and robustness in situations where other CDI methods fail. In a ptychographical microscope, the object is scanned area-by-area to record a diffraction pattern at each scan position, ensuring overlap between adjacent scan positions. The diffraction patterns and scan locations are then fed into an iterative algorithm, to retrieve the phase of the diffracted fields. By back-propagating the diffraction field to the sample position, complex-valued images of the object are obtained with quantitative amplitude and phase information. Scanning with overlap is a simple, yet elegant way of introducing redundant information into the diffraction-based measurements, distinguishing ptychography from other computational imaging techniques.

In addition to removing experimental restrictions required for other types of CDI, such as sample isolation, recent breakthroughs demonstrate that the information redundancy in ptychography also enables the reconstruction of state mixtures, i.e. the decomposition of mutually incoherent modes \cite{Batey2014, Thibault2013, Karl2015}. Thibault and Menzel \cite{Thibault2013} demonstrated the reconstruction of five dominant spatial modes in a partially coherent X-ray beam used for ptychography illumination. Shortly thereafter, Batey et al. \cite{Batey2014} simultaneously illuminated a sample with three separate single-color visible lasers with blue, green and red wavelengths, and successfully recovered the sample’s response for each of these three colors using the PIM technique.

CDI requires light sources with a high degree of spatial coherence. To address this challenge at short wavelengths, coherent light sources based on tabletop high harmonic generation (HHG) and large-scale free electron lasers (FELs) are under rapid development. High harmonic generation \cite{Rundquist1998, Bartels2000, Chen2010, Popmintchev2012, Popmintchev2015, Fan2015} is an extreme nonlinear process that coherently upconverts infrared, visible, or UV light to extreme ultraviolet (EUV) and X-ray wavelengths. HHG is unique as a light source because the generation conditions can be adjusted so that the light emerges either as a coherent supercontinuum (corresponding to an isolated attosecond burst in time), or as a comb of harmonics, with periodicity both in space and time. Under the right conditions, the harmonics are spectrally narrow ($\Delta\lambda/\lambda < 1 \%$) \cite{Bartels2000, Bartels2002OL} and also tunable by changing the driving laser wavelength or chirp. Since its demonstration in 2007, CDI using HHG sources has become a successful imaging modality, enabling high resolution image reconstruction both in transmission \cite{Sandberg2007, Seaberg2011, Zhang2013} and more recently in reflection \cite{Seaberg2014, Zhang2015, Gardner2012}. Despite these successes, HHG CDI contains untapped potential because the spectral structure of HHG has been underutilized. Indeed, recent work used broadband HHG beams as a way to make more efficient use of available photons \cite{Chen2009, Abbey2011}, by assuming that the object looks identical at all constituent wavelengths \cite{Chen2009, Abbey2011, Witte2014}. 

In this work, we extend PIM \cite{Batey2014} to utilize multiple high harmonics as the illumination source, in an important spectral range offering element-specific contrast due to adjacent absorption edges. By relaxing the previous assumption of a uniform spectral response, we retrieve a wealth of information about the sample, including the spectrally-dependent amplitude and phase that encode the sample composition and topography. To achieve this, we illuminate the sample with several harmonic orders simultaneously, and employ the PIM algorithm to reconstruct independent images at each wavelength. Thus, in addition to high-contrast, high-spatial-resolution imaging systems, we can now achieve intrinsic element-specific contrast at multiple wavelengths simultaneously. Information redundancy in ptychography not only allows the sample, but also the illumination to be reconstructed. Here we aslo demonstrate in situ imaging of a multicolor EUV beam. Finally, we note that the combination of a comb of coherent harmonics and the PIM algorithm is the most efficient use of HHG EUV radiation for imaging to date because there is no energy loss from any multilayer mirrors or monochromatizing optics. 

\section{Experiment}
The experimental setup is shown in Fig. 1. Phase matched high harmonics near 30 nm were generated by focusing ultrashort pulses from a Ti:Sapphire amplifier (KMLabs Dragon, 790 nm central wavelength, 5 kHz repetition rate, 1.5 mJ pulse energy, 22 fs pulse width) into a hollow waveguide (150 {\micro\meter} inner diameter, 5 cm long) filled with argon at 35 Torr. In contrast to previous experiments, no EUV multilayer mirrors were used to select a single harmonic. Instead, an ellipsoidal mirror refocused the all the harmonics onto the sample, with an estimated beam diameter of 10 {\micro\meter} based on a knife-edge measurement. 
Due to geometrical constraints of our imaging chamber, we added a pair of gold mirrors placed at an angle of 45{\degree} to steer the beam. 
The sample (titanium features patterned on silicon \cite{Seaberg2014}) was placed at a 50.5{\degree} angle of incidence. Diffraction from the sample, which is an incoherent superposition of scattered light from all four harmonics, was measured by an EUV-sensitive CCD (Andor) detector placed 55 mm away from the sample, and was positioned normal to the specular reflection of the beam. 
The ptychographical scan consisted of a 28 $\times$ 11 rectangular grid (with added random offsets up to 20\% of the step size to prevent periodic artifacts in the reconstruction). 
We chose a scanning step size of 1 {\micro\meter} to ensure enough information redundancy for the ptychographical dataset. The equivalent numerical aperture (NA) of the collected diffraction is 0.086, providing a half-pitch resolution near 170 nm.
The exposure time was set to 1.5 seconds to avoid saturation of the CCD. We accumulated three exposures at each position to increase the signal-to-noise ratio, resulting in a total exposure time of 23 minutes. In a chamber designed specifically for this experiment, the gold steering mirrors, which have a combined reflectivity of 2\%, could be removed and the required exposure time would then be reduced to 30 seconds using the HHG source described here.

\begin{figure}
\includegraphics[width=\textwidth]{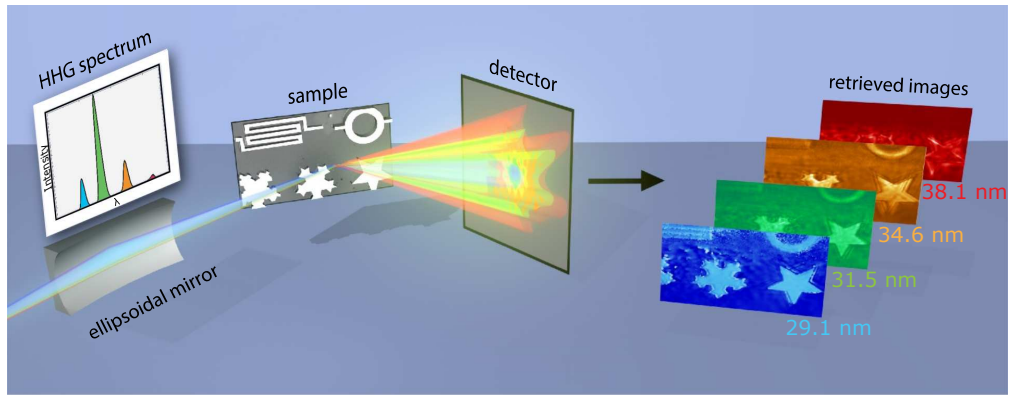} 
\caption{Hyperspectral imaging by combining multiple EUV harmonics and ptychographical information multiplexing. An EUV HHG beam consisting of four harmonic orders is focused onto the sample by a grazing-incidence ellipsoidal mirror. The sample is scanned transversely to enable ptychographical imaging. At each scan position a pixel-array detector records the diffraction pattern, which is an incoherent superposition of the four different wavelengths. The PIM algorithm can decompose the diffraction pattern into its coherent components, to obtain the spectral response of the sample at each wavelength simultaneously.}
\end{figure}

\begin{figure}
\includegraphics[width=\textwidth]{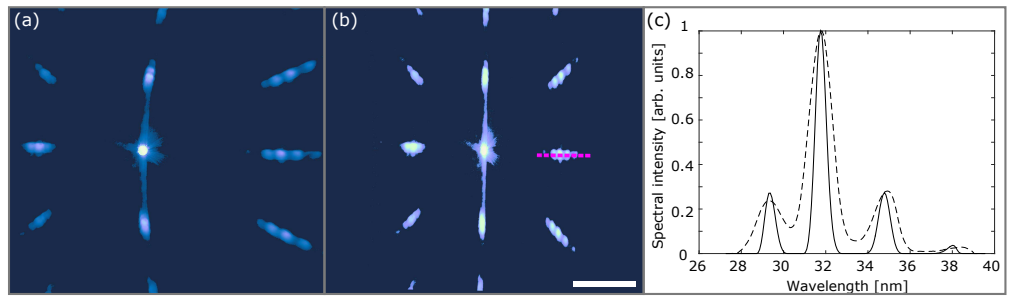} 
\caption{Spectrum of the multicolor EUV HHG source. (a) Diffraction intensity measured on the detector. (b) Diffraction amplitude obtained by performing tilted plane correction on (a). The axis is the spatial frequency normalized by $1/\lambda$, or effective NA (scale bar 0.02). (c) Spectral intensity. Dashed-line: line cut along the magenta line in (b). Solid line: higher-resolution estimate of the spectrum by deconvolving the modulus square of the remapped diffraction in (b).}
\end{figure}

To precisely determine the wavelengths of the harmonics, we placed a two-dimensional (2D), $\Lambda = 300$ nm period nano-pillar grating adjacent to the sample in the same plane. This grating acted as a low-resolution spectrometer. Diffraction from this grating illuminated by the EUV harmonic comb is shown in Fig.~2(a). We can clearly see the first-order diffraction peaks corresponding to spatial frequency $1/\Lambda$ for different harmonics. We remapped \cite{Gardner2012} the diffraction pattern to a uniform grid of spatial frequencies normalized by $1/\lambda$, with the result shown in Fig.~2(b). The spectral intensity, calculated as the modulus square of the diffraction amplitude along the magenta line in Fig.~2(b), is shown as a dashed line in Fig.~2(c), where the wavelength axis is calculated as the product of  and the normalized spatial frequency. The peaks are located at 29.1, 31.5, 34.6, and 38.1 nm, and correspond to harmonic orders 27, 25, 23, and 21 of the laser driving wavelength centered at 790 nm. The spectrum obtained in this way are blurred due to the fact that the HHG beam was not focused onto the detector (see dashed line in Fig.~2(c)). To obtain spectral intensities more representative of the actual HHG spectrum, we performed a 2D deconvolution of the remapped diffraction intensity using the Lucy-Richardson algorithm, with the zero-order diffraction taken to be the point spread function. The results of this analysis give an upper bound to the harmonic widths of $\Delta \lambda/\lambda \le  2\%$ (see solid line in Fig.~2(c)). We ignore the weak harmonic at 41 nm. The four-harmonic spectrum spans a width $\Delta \lambda/\lambda$ of 27\%. 

\section{Reconstruction}
We use a modified version of the PIM algorithm \cite{Batey2014} to reconstruct the response of the sample to the different HHG wavelengths present in the illumination. The modification is that we do not constrain the spectral weights during iterations, and let the relative scale of the probe and the sample float freely, while their product is invariant. This way, the algorithm will also work for applications where spectral weight information is missing, with the understanding that for each color, there is an undetermined relative scale factor between the reconstructed probe and the reconstructed object. The initial guess for the probe beams is obtained from a knife-edge focus measurement \cite{Zhang2015}. The entire reconstruction process consisted of $\approx$ 1300 iterations. During the final $\approx$ 100 iterations, we implemented the position refinement algorithm \cite{Zhang2013F} for each color.

\begin{figure}
\includegraphics[width=\textwidth]{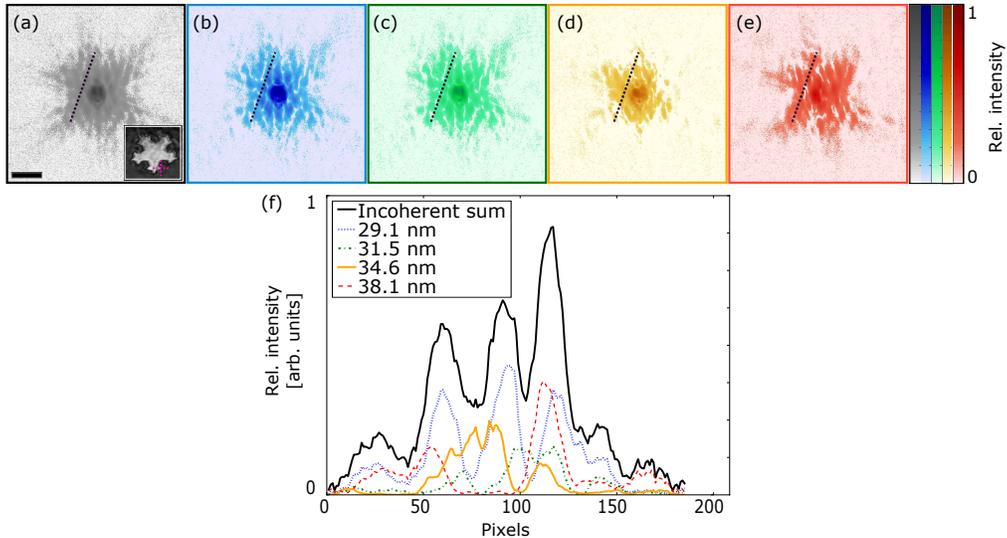} 
\caption{Decomposition of a multicolor diffraction pattern. (a) Measured diffraction intensity after remapping around the position marked by magenta crosshair in the inset. (b-e) The coherent components of (a) at 29.1 nm, 31.5 nm, 34.6 nm, and 38.1 nm respectively. All the diffraction intensities in (a-e) are shown to the quarter power, and share the scale bar with (a), which has a length of 0.02 NA. (f) Comparison of the intensity profile of the incoherent superposition in (a) and its components in (b-e) along the dashed-line.}
\end{figure}

An illustration of how multicolor diffraction patterns are decomposed using PIM is shown in Fig.~3. We select the scan position centered around the crosshair mark in the inset of Fig.~3(a), and display the measured diffraction intensity (after remapping \cite{Gardner2012}). Figure~3(b-e) show the four quasi-monochromatic components, as determined by the PIM algorithm, that sum incoherently to form the pattern shown in Fig.~3(a). Figure~3(f) shows the intensity profiles of Fig.~3(a-e) along the dashed line. The lower contrast of the diffraction fringes in the mixed diffraction pattern in Fig.~3(a) relative to that of its coherent components is clear. The different colors have different scattering angles for the same spatial frequencies and different spatial distribution of illumination fields, causing blurring in the mixed diffraction pattern. 

\begin{figure}
\includegraphics[width=\textwidth]{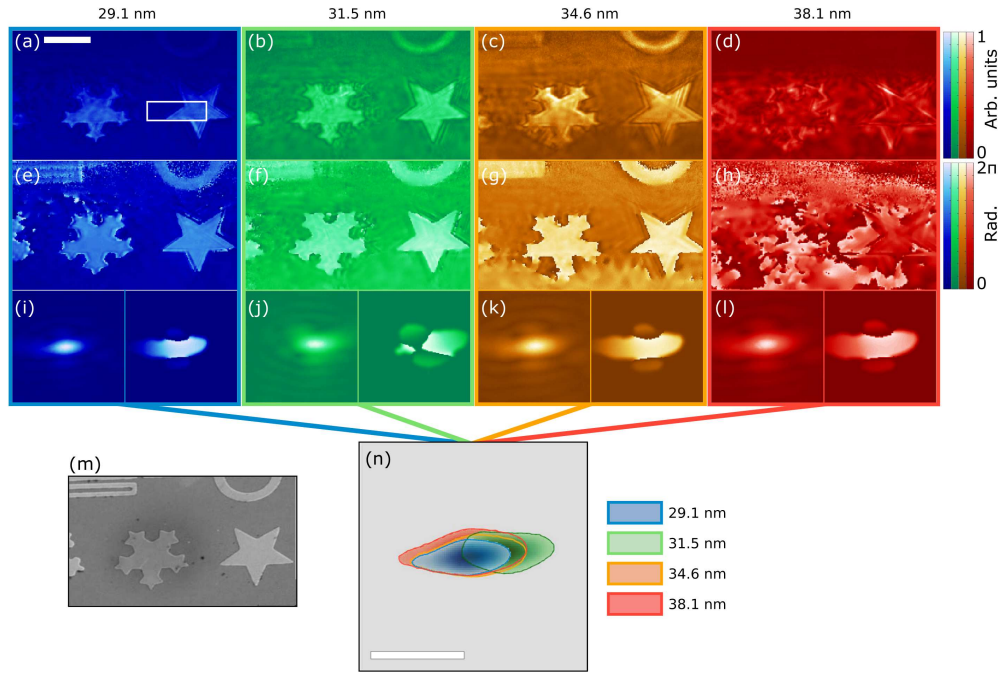} 
\caption{Reconstructed sample reflection coefficients for the four harmonics in the illumination. The first row (a-d) shows the amplitudes and the second row (e-h) shows the phases. The third row (i-l) shows the normalized amplitude (left) and the phase (right) of the probe for each color. Each color has the amplitude normalized so that its maximum is one. (m) For comparison, a scanning electron microscopy image of the sample is shown. (n) The combination of probes at four different wavelengths to show the spatial dispersion of the HHG beam. Scale bar in (a) and (n), 10 {\micro\meter}. (b-m) share the same scale bar as in (a).  The white box in (a) marks a region of interest for later quantitative analysis.}
\end{figure}

We reconstructed the complex amplitude of the reflection from the sample, with the moduli shown in Fig.~4(a-d) and phases shown in Fig.~4(e-h). As a reference, an SEM image of the sample is shown in Fig.~4(m). Excluding the reconstruction for the 21st harmonic, the main features of the sample are reconstructed well. There are several types of artifacts present in the PIM reconstructions, such as double-lining near edges and noise on the silicon substrate. Compared with our previous reconstruction with single color illumination \cite{Zhang2015}, the multicolor reconstruction has degraded quality. The probable reason is that while the signal-to-noise ratio for the sum of all four colors is about the same as for the single-color case, after decomposition into four colors, each color has much lower signal-to-noise ratio leading to lower image quality. This is mainly a technical limitation due to the limited dynamic range of the detector used in this experiment, and should not be considered to be a fundamental limitation to the technique. The image reconstruction quality is especially low for the 21st harmonic, which not only shows a low spectral weight in the measured spectrum in Fig.~2(c) but also has a lower reflectivity from titanium: the reflectivity at 38.1 nm is only 11\% of that at 29.1 nm \cite{Henke1993}. Ignoring other weak harmonics present in the spectrum may also lead to reconstruction errors. 

Reconstructed moduli and phases of the HHG beam at four different wavelengths are shown in Fig.~4(i-l), with Fig.~4(n) showing the different color beams on the same plane to provide a direct view of the entire HHG beam. The outline of each beam in Fig.~4(n) corresponds to the $1/e^2$ intensity level. In terms of the HHG probe beam reconstruction, we can clearly see astigmatism in Fig.~4(i-l), with the wavefront converging horizontally and diverging vertically. This extracted wavefront curvature agrees with our observations during the experiment: using a knife-edge measurement we found that the x-focus is after the y-focus by roughly 400 {\micro\meter} along the beam axis, and we positioned the sample at the midpoint of the x- and y-foci.

\begin{figure}
\includegraphics[width=\textwidth]{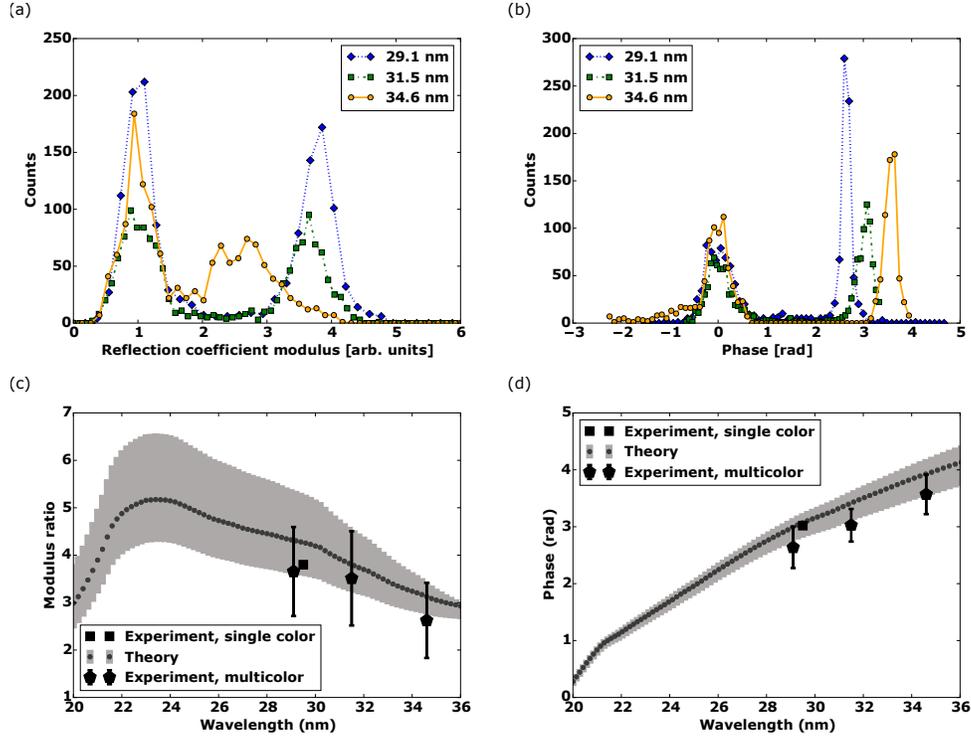} 
\caption{Spectral contrast analysis. For the area enclosed by the white box in Fig.~4(a), a histogram of the modulus and the phase of the reflection from the sample is shown in (a) and (b) respectively. For convenience, in (a) the reflection coefficient moduli $|r|$ are normalized such that $|r|=1$ for silicon for all wavelengths; while in (b) the phases for silicon are set to zero for all wavelengths. (c) and (d) show the comparison of modulus $|r_{\mathrm Ti} |/|r_{\mathrm Si} |$ and phase difference $\phi_{\mathrm Ti}-\phi_{\mathrm Si}$ from theory, the single color experiment in previous work, and the multicolor experiment in this work.}
\end{figure}

\section{Spectral contrast analysis} 
To verify the spectral contrast obtained in the PIM experiments, we performed a comparison against both the previous single-color work \cite{Seaberg2014} as well as theoretical calculations. For the region marked by a white rectangle in Fig.~4(a), we performed histogram analysis of the modulus ratio of the titanium region versus that of the silicon region $|r_{\mathrm Ti}  |/|r_{\mathrm Si}|$, and also the phase difference between the two regions, $\phi_{\mathrm Ti}-\phi_{\mathrm Si}$, as shown in Fig.~5(a) and Fig.~5(b).  To obtain the theoretical values of these two quantities, we used the results from a previous characterization of the same sample \cite{Seaberg2014}: the thickness of the oxidation layers on top of the titanium and the silicon regions are assumed to be 2.92 nm and 3.01 nm thick, respectively. From an atomic force microscopy measurement, we characterized the roughness of the titanium region and the silicon region to be 0.91 nm and 0.23 nm respectively, and the height difference $h_{\mathrm Ti} -h_{\mathrm Si}$  to be 32.7 nm. We found evidence of up to 1 nm of carbon buildup in samples maintained in the same environment as the one used in this work, using Auger electron spectroscopy. Here we assume a thickness between 0 and 1 nm of the carbon contamination, which is a source of uncertainty in the theoretical reflectivities of the sample surfaces. With this model of the sample, and tabulated values of indices of refractions and reflectivities \cite{Henke1993}, we calculated $|r_{\mathrm Ti} |/|r_{\mathrm Si} |$ and $\phi_{\mathrm Ti}-\phi_{\mathrm Si}$ (the total phase results from both the complex Fresnel reflection coefficient and the height difference \cite{Seaberg2014, Zhang2015}) as a function of wavelength. The comparison between these calculations and the results from the previous single color experiment \cite{Seaberg2014} along with the current multicolor experiment is shown in Fig.~5(c) and Fig.~5(d). Both of these comparisons demonstrate good agreement between the multicolor and single-color experimental results and theoretical simulations. 

\section{Discussions and conclusions}

As illustrated above, the combination of high harmonic combs and ptychographic information multiplexed diffraction imaging allows the amplitude and phase response of a sample to be recovered at multiple wavelengths simultaneously. Each of the four illuminating probes, one for each phase-matched harmonic, is reconstructed separately, over a spectral range corresponding to $\Delta\lambda/\lambda \approx 27\%$. No wavelength-scanning or separation hardware was used, making the experimental setup simpler than other techniques while at the same time enabling more efficient use of available photons. The combination of a comb of coherent harmonics with the PIM algorithm is the most efficient use of HHG EUV radiation for imaging to date because there is no energy loss from any multilayer mirrors or monochromatizing optics.

The limitations to this new spectromicroscopy technique are not currently well understood, including how many colors, $N_{\mathrm c}$, can be reconstructed simultaneously, or what spectral resolving power, $\lambda/\Delta\lambda$, can be achieved. The limit for $N_{\mathrm c}$ can be estimated by simply considering the ratio between the number of knowns and unknowns in the phase retrieval problem. If a point on the sample is illuminated with $N_{\mathrm p}$ overlapping ptychography scan positions, i.e. it is independently measured $N_{\mathrm p}$ times, and the diffraction pattern is sampled at an oversampling \cite{Bates1982} ratio $\rho$, then there is $N_{\mathrm p} \rho$-fold information redundancy for that point. Beyond $\rho=2$, no further information is gained, so for $\rho>2$ the information redundancy is simply $2N_{\mathrm p}$. Each color needs two-fold redundancy for phase retrieval, so an upper bound for $N_{\mathrm c}$ is $N_{\mathrm p} \rho/2$ for $\rho\le2$, and $N_\mathrm{p}$ for $\rho>2$. It is reasonable to believe that there is a point beyond which increasing $N_{\mathrm p}$ (increasing overlap between scan positions) fails to result in increased information redundancy i.e. adjacent scans are no longer independent. Regarding spectral resolution, the detection of scattered light at high numerical aperture can help separate adjacent colors via diffraction. In object space, this translates to the fact that in order to discriminate between adjacent colors, the respective fields of view must differ in image size (number of pixels). 
With this in mind the spectral resolution is limited to approximately $\lambda/\Delta\lambda \ge 2 \mathrm{NA} L/\lambda$, where $L$ is the maximum distance between scan positions, and NA is the numerical aperture. Thus an increase in either the NA or the maximum distance between scan positions can result in increased spectral resolution. For the experiment described here, the estimated upper bound for $N_c$ is $\approx$100, much greater than the 4 colors reconstructed here. The spectral resolving power is estimated to be 80, which is higher than the necessary resolving power of 15 necessary to distinguish adjacent harmonics.

Looking forward, the spatial resolution can easily reach below 50 nm by placing the detector closer to the sample for a higher numerical aperture \cite{Zhang2015}. Various element-specific absorption edges in the EUV/X-ray spectrum range at higher photon energies than studied here provide opportunities for high-contrast spectral imaging. Broadband and narrowband shorter wavelength HHG sources are being developed that are bright (i.e. phase matched) and that can operate in the water window range \cite{Chen2010, Popmintchev2015} for biological imaging, reaching photon energies up to 1.6 keV \cite{Popmintchev2012}. It is also possible to extend this technique to multicolor sources other than HHG, such as synchrotrons and free electron lasers. The HHG source used in this work has a favorable spectral structure with discrete spectral lines, each of which has a narrow bandwidth. The question of how to address more continuous, broadband spectra with this technique will be addressed in future work. Each harmonic has a transform limited pulse duration on the order of $\approx 5$ femtoseconds (30 nm wavelength with a spectral bandwidth of $\Delta\lambda/\lambda = 1\%$ and a Gaussian profile corresponds to pulse duration of 5 femtoseconds). By making use of the ultrashort pulse duration of the HHG source in a pump-probe geometry, this technique can be used for spectral imaging of ultrafast charge, spin and phonon dynamics in functioning nanosystems \cite{Hoogeboom-Pot2015, Turgut2013}.
\section*{Acknowledgements}
	This work was performed at JILA. We thank Dr. Ming-Chang Chen for his help on the ellipsoidal mirror. We also gratefully acknowledge support from the Defense Advanced Research Projects Agency (DARPA) PULSE program, the NSF Engineering Center in EUV Science and Technology, the Semiconductor Research Corporation (2013-OJ-2443), the National Science Foundation COSI IGERT 0801680, the National Science Foundation Graduate Research Fellowship Program, Ford Foundation fellowship program, and the Katherine Burr Blodgett fellowship program. The current address of B.Z. is KLA-Tencor Corporation, One Technology Drive, Milpitas, CA 95035, USA.  The current address of M.H.S. is SLAC National Accelerator Laboratory, Menlo Park, CA 94025, USA. 
 
\end{document}